\documentclass[
reprint,
 amsmath,amssymb,
 aps,
showkeys,
showpacs
]{revtex4-2}

\usepackage{comment}
\usepackage{graphicx}
\usepackage{dcolumn}
\usepackage{bm}
\usepackage{slashed}
\usepackage{mathrsfs}

\newcommand{\bone}{1\!\!1}
\newcommand{\e}{\textrm{e}}

\begin{document}

\preprint{MUPB/Conference section: }

\title{Obtaining fully polarised amplitudes in gauge invariant form\\}

\author{Naser Ahmadiniaz}
\email{n.ahmadiniaz@hzdr.de}
\affiliation{Helmholtz-Zentrum Dresden-Rossendorf, Bautzner Landstraße 400, 01328 Dresden, Germany}
 
\author{Victor Miguel Banda Guzman}
\email{victor.banda@umich.mx}
\affiliation{Instituto de Física y Matemáticas,\\
Universidad Michoacana de San Nicolás de Hidalgo,  Morelia, México}

\author{Fiorenzo Bastianelli}
\email{bastianelli@bo.infn.it}
\affiliation{Dipartimento di Fisica e Astronomia “Augusto Righi”, Università di Bologna, Via Irnerio 46,
I-40126 Bologna, Italy}

\author{Olindo Corradini}
\email{olindo.corradini@unimore.it}
\affiliation{Dipartimento di Scienze Fisiche, Informatiche e Matematiche, Università degli Studi di Modena e Reggio Emilia, Via Campi 213/A, I-41125 Modena, Italy}

\author{Christian Schubert}
\email{christianschubert137@gmail.com}
\affiliation{Instituto de Física y Matemáticas,\\
Universidad Michoacana de San Nicolás de Hidalgo,  Morelia, México}

\author{\underline{James P. Edwards}}
\affiliation{Present address, Centre for Mathematical Sciences, University of Plymouth, Plymouth, PL4 8AA, UK.}
\email{jpedwards@cantab.net (Corresponding author)}

\date{\today}

\begin{abstract}
We describe progress applying the \textit{Worldline Formalism} of quantum field theory to the fermion propagator dressed by $N$-photons to study multi-linear Compton scattering processes, explaining how this approach -- whose calculational advantages are well-known at multi-loop order -- yields compact and manifestly gauge invariant scattering amplitudes.
\end{abstract}

\maketitle


\section{Introduction}
\label{secIntro}
Quantum field theory has a long history and perturbative calculation of scattering amplitudes in the standard Feynman diagram formalism is well established. Yet in the same papers that developed this approach \cite{Feynman:1950ir, Feynman:1951gn}, Feynman hinted at an alternative, \textit{first quantised} formulation of field theory that later evolved into the \textit{worldline formalism} \cite{Strass1} -- for reviews see \cite{ChrisRev, UsRep}. Historical applications within this alternative framework focussed on calculating the effective action \cite{Strass1} or multi-loop amplitudes \cite{Schmidt:1994aq} and the worldline approach was only recently extended to type of tree-level amplitudes we highlight here \cite{fppaper1, fppaper2}. In this contribution we report the salient formulae describing the fermion propagator dressed by $N$-photons and describe how to extract manifestly gauge invariant amplitudes for multi-photon emission or absorption from the line. 

We organise the presentation by starting with the path integral representation of the dressed propagator in section \ref{secProp}. We follow this in section \ref{secAmps} with the study of photon amplitudes. We then conclude with an outlook for future work in this area.

\section{Dressed propagator}
\label{secProp}
As is well known, the perturbative calculation of scattering amplitudes is met with the problems of rapid growth in both the number and complexity of Feynman diagrams (well-illustrated by calculations of the electron anomalous magnetic moment). The worldline formalism naturally sums over diagrams related by e.g. exchange of external legs, making it a good candidate for overcoming these issues. Working within quantum electrodynamics, the worldline representation of the electron propagator in an electromagnetic background $A(x)$, defined by ${S^{x'x} = \big\langle x' \big|\big[m - i\slashed{D} \big]^{-1} \big| x\big\rangle}$, is found by exploiting the Gordon identity as in \cite{fppaper1}, by writing
\begin{align}
	 S^{x'x}	&= \big[m + i\slashed{D}' \big]\big\langle x' \big|\big[m^2 - D^2 + \frac{ie}{2} \gamma^{\mu}F_{\mu\nu}\gamma^{\nu}\big]^{-1} \big| x\big\rangle \nonumber \\
	 & \equiv \big[m + i\slashed{D}' \big]K^{x'x}[A]\,,
\end{align}
where $D'_{\mu} = \partial^{\prime}_{\mu} + ieA_{\mu}$, and writing the second order formalism kernel, $K^{x'x}$, as the following path integral
\begin{equation}
	\hspace{-0.75em}K^{x'x} = 2^{-\frac{D}{2}}\textrm{symb}^{-1}  \int_{0}^{\infty}dT\, \e^{-m^{2}T}\int_{x(0) = x}^{x(T) = x'}\hspace{-1.25em} \mathscr{D}x \int_{APC}\hspace{-0.5em}\mathscr{D}\psi \, \e^{-S[x, \psi]}\,,
\end{equation}
over auxiliary point particle trajectories propagating from $x$ to $x'$ in proper time $T$ with Grassmann fields $\psi(x)$ defined along them, whose dynamics are specified by the worldline action
\begin{align}
	S\left[x, \psi\right] = \int_{0}^{T}d\tau \big[ \frac{\dot{x}^{2}}{4} &+ \frac{1}{2}\psi \cdot \dot{\psi} + ie A(x) \cdot \dot{x} \nonumber \\
	&- ie(\psi + \eta)^{\mu}F_{\mu\nu}(x) (\psi + \eta)^{\nu} \big]\,.
\end{align}
The \textit{symbol map} acts on the constant Grassmann vectors to produce the $\gamma$-matrix structure of the propagator according to $\textrm{symb}(\gamma^{\alpha_{1}\ldots \alpha_{n}}) = (-i\sqrt{2})^{n}\eta^{\alpha_{1}}\cdots\eta^{\alpha_{n}}$, where $\gamma^{\alpha_{1}\ldots \alpha_{n}}$ is the antisymmetrised product of $n$ matrices $\gamma^{\alpha_{i}}$. So far this treatment is valid for an arbitrary background electromagnetic field -- we now show how to obtain photon scattering amplitudes from this dressed propagator. 

\vspace{-1em}
\section{Photon amplitudes}
\label{secAmps}
\begin{figure}
	\centering
	\includegraphics[width=0.45\textwidth]{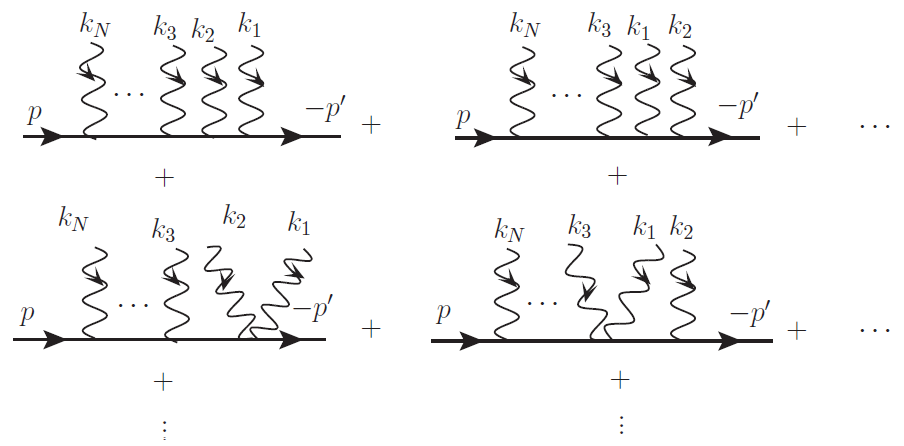}
	\caption{The $N$-photon amplitudes generated by (\ref{eqSLeadSub}) -- note that the sum over permutations of diagrams related by exchanging external photon legs is automatically included.}
	\label{figNPhoton}
\end{figure}

Extracting photon amplitudes requires fixing the background field to a sum of plane wave asymptotic states, 
\begin{equation}
A^{\mu}(x) = \sum_	{i=0}^{N} \varepsilon_{i}^{\mu} \e^{i k_{i} \cdot x}\,,
\end{equation}
which leads to interactions with photons represented by \textit{vertex operators} under the path integral, written as 
\begin{equation}
	V_{\eta}^{x'x}[k, \varepsilon] = \int_{0}^{T}d\tau\, \big[ \varepsilon\cdot \dot{x} + 2ie\varepsilon \cdot (\psi + \eta)k\cdot (\psi + \eta) \big] \e^{i k \cdot x}\,,
\end{equation}
similarly to as in string theory. Transforming to momentum space the dressed propagator is divided into ``\textit{leading}'' and ``subleading'' terms according to (the hats indicate variables to be removed from the argument)
\begin{align}
	&S^{p'p}_{N}[k_{1}, \varepsilon_{1}; \ldots ;k_{N}, \varepsilon_{N}] = (\slashed{p}' + m)K_{N}^{p'p}[k_{1}, \varepsilon_{1}; \ldots ; k_{N}, \varepsilon_{N}]\nonumber \\
	&- e\sum_{j=1}^{N}\slashed{\varepsilon}_{j}K_{N-1}^{p'+k_{j},p}[k_{1}, \varepsilon_{1}; \ldots;\hat{k}_{j},\hat{\varepsilon}_{j}; \ldots; k_{N}, \varepsilon_{N}]\,.
	\label{eqSLeadSub}
\end{align}
The amplitudes generated by this propagator are illustrated in figure \ref{figNPhoton}.

\subsection{Gauge Invariance}
For an on-shell process, the amplitude must satisfy the Ward identity. This can be checked by verifying its invariance under the transformation $\varepsilon_{i} \rightarrow \varepsilon_{i} + \lambda_{i} k_{i}$ for constant $\lambda_{i}$, which usually requires cancellation between contributions from various Feynman diagrams. Since the worldline formalism automatically sums over diagrams related by exchange of external legs we can hope to manifest its transversality already at the level of the integrand. Indeed this can be achieved if we rewrite the vertex operator by introducing a total derivative as \cite{fppaper2} 
\begin{align}
	\hspace{-1.2em}V_{\eta}^{x'x}[k, \varepsilon] &= \int_{0}^{T}d\tau\, \big[ \varepsilon\cdot \dot{x} + i\frac{\varepsilon\cdot r}{k\cdot r} \frac{d}{d\tau} - ie(\psi + \eta)\cdot f \cdot (\psi + \eta) \big] \e^{i k \cdot x} \\
	\hspace{-1.6em}&= \int_{0}^{T}d\tau\, \big[ \frac{r\cdot  f \cdot \dot{x}}{r\cdot k} - ie(\psi + \eta)\cdot f \cdot (\psi + \eta) \big] \e^{i k \cdot x}\,
\end{align}
involving an auxiliary reference vector $r$ -- this can be chosen arbitrarily as long as $r \cdot k \neq 0$. 

Now, both the boundary terms introduced in the new vertex operator and the subleading terms in (\ref{eqSLeadSub}) vanish when the LSZ prescription is applied to amputate external fermion legs since they are missing the appropriate pole structure, so we arrive at the amputated amplitude
\begin{align}
	\mathcal{M}^{p'p}_{N} &= \bar{u}(-p')(-\slashed{p}^{\prime} + m)(\slashed{p}^{\prime} + m)S_{N}^{p'p}(\slashed{p} + m)u(p) \nonumber \\
	&= \bar{u}(-p') \frac{\mathfrak{K}_{N}^{p'p}}{2m}u(p)\,,
\end{align}
where $\mathfrak{K}_{N}$ is the on-shell residue of $K_{N}$,
\begin{equation}
	K_{N}^{p'p} \equiv (-ie)^{N} \frac{\mathfrak{K}_{N}^{p'p}}{(p'^{2} + m^{2})(p^{2} + m^{2})}\,,
\end{equation} 
that is now a function of the transverse field strength tensors, $f_{i}$. Hence we have arrived at a clearly gauge invariant formula for on-shell amplitudes. The experienced reader will recognise the potential difficulties in rewriting a general amplitude -- even one as simple as the Compton amplitude -- expressed as a function of photon polarisations in terms of their field strength tensors.

\vspace{-1.5em}
\subsection{Applications}
To illustrate this approach we recall some simple examples from \cite{fppaper1, fppaper2}. To begin, note that the symbol map implies the kernel is naturally written in terms of elements of the even sub-algebra of the Dirac representation of the Clifford algebra, so we can decompose ($\sigma^{\mu\nu} = \frac{1}{2}[\gamma^{\mu}, \gamma^{\nu}]$)
\begin{equation}
	 \mathfrak{K}_{N}^{p'p} \equiv A_{N}\bone + B_{N\alpha\beta}\sigma^{\alpha\beta} -iC_{N}\gamma_{5}\,.
\end{equation}
The simplest example is with $N = 0$, for which $A_{0} = 1$ and $B_{0} = 0 = C_{0}$ which lead to $\mathfrak{K}_{0}^{p'p} = 1$ and
\begin{align}
\hspace{-1.25em}	K_{0}^{p'p} = \frac{1}{(p'^{2} + m^{2})(p^{2} + m^{2})} \Longrightarrow  S_{0}^{p'p} &= \frac{\slashed{p}^{\prime} + m}{(p'^{2} + m^{2})(p^{2} + m^{2})}\,,
\end{align}
which confirms the bare propagator is correctly obtained (un-amputated with respect to external fermion legs).

The vertex is reproduced for $N = 1$, whereby $C_{1} = 0$ and the non-zero coefficients are
\begin{equation}
	A_{1} = -2\frac{r\cdot f \cdot p}{r\cdot k} \,, \qquad B_{1\alpha\beta} = \frac{1}{2}f_{\alpha\beta}\,.
\end{equation}
Expanding the field strength tensor, the first of these can equally be written on-shell as $A_{1} = -2\varepsilon \cdot p = -\varepsilon \cdot (p - p')$, which is transverse thanks to momentum conservation. A quick calculation then verifies that
\begin{align}
	\hspace{-0.5em}K_{1}^{p'p} = \frac{\varepsilon \cdot (p' - p) \bone + \frac{1}{2} (\slashed{k} \slashed{\varepsilon} - \slashed{\varepsilon} \slashed{k})}{(p'^{2} + m^{2})(p^{2} + m^{2})} = \frac{\varepsilon (\slashed{p} - m) - (\slashed{p}^{\prime} - m)\slashed{\varepsilon}}{(p'^{2} + m^{2})(p^{2} + m^{2})} \,,
\end{align}
so that
\begin{equation}
	S_{1}^{p'p} = \frac{(\slashed{p}^{\prime} + m)\slashed{\varepsilon} (\slashed{p} - m)}{(p'^{2} + m^{2})(p^{2} + m^{2})}\,,
\end{equation}
easily recognised as the un-amputated Dirac vertex.

The Compton amplitude with $N = 2$ photons requires determination of the following coefficients:
\begin{align}
	\hspace{-0.25em}A_{2} &=-2\frac{r_{1}\cdot f_{1} \cdot f_{2} \cdot r_{2}}{r_{1}\cdot k_{1} r_{2}\cdot k_{2}}- \frac{1}{2}\Big[\frac{1}{2p' \cdot k_{1}} + \frac{1}{2p'\cdot k_{2}}\Big] \textrm{tr}\big(f_{1} \cdot f_{2}\big)	\nonumber \\
	\hspace{-0.25em}B_{2}^{\alpha\beta} &= -\frac{1}{2} \frac{r_{1}\cdot f_{1} \cdot k_{2} f_{2}^{\alpha\beta} + r_{2}\cdot f_{2} \cdot k_{1} f_{1}^{\alpha \beta}}{r_{1}  \cdot k_{1} r_{2} \cdot k_{2}} \nonumber \\
	\hspace{-0.25em}&+ \frac{1}{2}\Big[\frac{1}{2p' \cdot k_{1}} - \frac{1}{2p'\cdot k_{2}}\Big] \big[f_{1}, f_{2}\big]^{\alpha\beta} \nonumber \\
	\hspace{-0.25em}C_{2} &= \frac{1}{2}\Big[\frac{1}{2p' \cdot k_{1}} + \frac{1}{2p'\cdot k_{2}}\Big] \textrm{tr}\big(f_{1} \cdot \widetilde{f}_{2}\big)\,,
\end{align}
where $\tilde{f}$ is the dual field strength tensor. Useful choices for the reference vectors include $r_{i} = p^{\prime}$ and $r_{i} = p^{\prime} + \frac{m^{2}}{2p' \cdot k_{i}}k_{i}$ -- see \cite{fppaper2} for more details and identities allowing for $A$ and $C$ to be constructed from $B_{\mu\nu}$ on-shell. 

In \cite{fppaper2} these are used to construct the fully polarised amplitude, the spin-summed amplitude with fixed photon helicities and the un-polarised amplitude. Indeed, there a particularly simple ``diagonal'' formula was found for the spin / polarisation summed amplitudes:
\begin{equation}
	\big\langle\left|\mathcal{M}_{N}\right|^{2}\big\rangle = e^{2N}\left[ \left|A_{N}\right|^2 + 2B_{N}^{\alpha\beta}B^{\star}_{N\alpha\beta} - \left|C_{N}\right|^{2}\right]\,,
\end{equation}
that bypasses the need to construct an explicit expression for the amplitude itself. The same formalism can also be used to generate higher loop contributions to the scattering processes by sewing pairs of photons. This was shown in \cite{fppaper1} for the case of the one-loop self-energy.

\hspace{1.5em}

\section{Conclusion}
We have summarised the worldline description of the fermion propagator and multi-linear Compton scattering, arriving at a compact form of the amplitude in terms of manifestly gauge invariant photon field strength tensors. Its advantages include the early projection onto antisymmetrised products of $\gamma$-matrices that avoids traces of long products of them that appear in the standard approach, and which also allows summing over fermion polarisations without fixing the number or helicities 
of the photons; and a natural ``spin-orbit decomposition'' that is presented in \cite{fppaper1}. The new approach has also exposed  previously unknown relations between amplitudes in scalar and spinor QED and recursion relations between the coefficients of $\mathfrak{K}_{N}$ for different numbers of photons \cite{fppaper2}. 

Ongoing work applies these tools to analyse the gauge properties of the propagator at higher loop order (via the LKF transformations \cite{UsLKFT}) reported on by J. Nicasio \cite{ProcNicasio} elsewhere in these proceedings and extending the formalism to describe scattering amplitudes in the presence of constant and plane wave electromagnetic backgrounds. 


%

\bibliography{bibLomCon}

\end{document}